\documentclass[spanish,american,pre, twocolumn]{revtex4}
\usepackage[T1]{fontenc}
\usepackage[latin9]{inputenc}
\usepackage{amsmath}
\usepackage{amssymb}
\usepackage{graphicx}

\makeatletter

\providecommand{\tabularnewline}{\\}
\newcommand{\lyxdot}{.}

\@ifundefined{textcolor}{}
{%
 \definecolor{BLACK}{gray}{0}
 \definecolor{WHITE}{gray}{1}
 \definecolor{RED}{rgb}{1,0,0}
 \definecolor{GREEN}{rgb}{0,1,0}
 \definecolor{BLUE}{rgb}{0,0,1}
 \definecolor{CYAN}{cmyk}{1,0,0,0}
 \definecolor{MAGENTA}{cmyk}{0,1,0,0}
 \definecolor{YELLOW}{cmyk}{0,0,1,0}
}

\makeatother

\usepackage{babel}
\addto\shorthandsspanish{\spanishdeactivate{~<>}}

\begin{document}

\title{Fluctuation-induced forces between rings threaded around a polymer chain
under tension}

\author{F. M. Gilles$^{1,2,3\dagger}$, R. Llubaroff$^{1,4\dagger}$ and
C. Pastorino$^{1,2}$}

\affiliation{$^{1}$Departamento de Física de la Materia Condensada, CAC-CNEA,
Av.Gral.~Paz 1499, 1650, Pcia.~de Buenos Aires, Argentina}
\email{pastor@cnea.gov.ar}

\affiliation{$^{2}$ CONICET, Godoy Cruz 2290 (C1425FQB), Buenos Aires, Argentina}

\affiliation{$^{3}$ Instituto de Investigaciones Fisicoquímicas Teóricas y Aplicadas (INIFTA), 
Departamento de Química, Facultad de Ciencias Exactas, Universidad Nacional de La Plata, La Plata 1900, Argentina
}

\affiliation{$^{4}$Facultad Regional Avellaneda, Universidad Tecnológica Nacional
(UTN-FRA)\\
$^{\dagger}$ F. M. Gilles and R. Llubaroff contributed to this work
on an equal basis.}
\begin{abstract}
We characterize the fluctuation properties of a polymer chain under
external tension and the fluctuation-induced forces between two ring
molecules threaded around the chain. The problem is relevant in the context
of fluctuation-induced forces in soft matter systems, features of
liquid interfaces and to describe properties of polyrotaxanes and
slide-ring materials. We perform molecular dynamics simulations of
the Kremer-Grest bead-spring model for the polymer and a simple ring-molecule
model, in the canonical ensemble. We study transverse fluctuations
of the stretched chain, as a function of chain stretching and in the
presence of ring-shaped threaded molecules. The fluctuation spectra
of the chains are analyzed in equilibrium at constant temperature
and the differences in presence of two ring molecules are compared.
For the rings located at fixed distances, we find an attractive fluctuation-induced
force between the rings, proportional to the temperature and decaying
with the ring distance. We characterize this force as a function of
ring distance, chain stretching, ring radius and measure the differences
between the free chain spectrum and the fluctuations of the chain
constrained by the rings. We also compare the dependence and range
of the force found in the simulations with theoretical models coming
from different fields. 
\end{abstract}
\maketitle

\section{Introduction}

Fluctuation-induced forces have attracted enormous attention starting
from the renowned Casimir effect, which was discovered in the context
of quantum electrodynamics\cite{Casimir_48}. However, the key elements
for the existence of fluctuation-induced forces are present in a broad
range of systems. Those elements are a fluctuating medium and an external
object, whose presence inhibits or hinders the natural fluctuations
of the medium\cite{Kardar_99}. The first realm of study of fluctuation-induced
forces was given by quantum fluctuations of the electromagnetic field
with restrictions imposed by perfect parallel conducting plates, as
in the seminal work by Casimir\cite{Casimir_48,Dalvit_11}. However,
forces arising from thermal fluctuations of the electromagnetic field
have also been predicted and measured experimentally\cite{Sushkov_11}.
A variety of physical systems were found to present effective forces
of the same origin due to thermal fluctuations of material fields
and molecules. For example, colloids located at liquid-liquid interfaces
act as physical restrictions of interface fluctuations. Fisher and
De Gennes noted that these interactions should appear in a film of
a binary liquid mixture close to a wall at the critical point, where
the correlation of concentration fluctuations diverges\cite{Fisher_78,Kardar_99}.
The objects restricting fluctuations are in this case the interface
with the wall and the gas-liquid film interface, which are affected
by an attractive fluctuation-induced force. This force, known as critical
Casimir effect, was measured recently using colloids and total reflection
microscopy\cite{Hertlein_08,Gambassi_09a}. Beyond being a fascinating
physical effect\cite{Lamoureaux_07}, the fluctuation-induced forces
became relevant in practice due to miniaturization and manipulation
of matter at the nanoscale. This ranges from the development of micro
and nano electromechanical systems (MEMS)\cite{Dalvit_11} and the
behavior of colloids or proteins in interfaces and membranes\cite{Lehle_06,Bresme_09,Kardar_99,Weikl_01,Machta_12}.
The range of the force is related to the characteristic length of
correlation of fluctuations, which becomes comparable with mesoscopic
distances between colloids, molecules or aggregates in many systems.
Fluctuation-induced forces have been studied in superfluid films\cite{Ganshin_06,Garcia_99}
, liquid crystals, inclusions or proteins in membranes\cite{Reynwar_08,Weikl_01,Lin_11}
and colloids confined in liquid interfaces\cite{Lehle_06,Bresme_09,Noruzifar_09,Noruzifar_13,Kardar_99}.
More recently, Casimir-like forces were studied in out-of equilibrium
diffusive systems\cite{Aminov_15} and active matter\cite{Ray_14,Parra-Rojas_14}.

Within the context of soft matter, we study the fluctuation-induced
forces between ring molecules threaded around a polymer chain under tension
at thermal equilibrium. The system is interesting from a basic point
of view, but can be also synthesized in the form of supramolecular
aggregates, the so-called polyrotaxanes. A polyrotaxane is formed
by a varying number of ring molecules, usually cyclodextrin, threaded
in a backbone linear polymer chain.\cite{Araki_07} The end-groups
of the polymer are big enough such that the rings cannot get out of
the chain, or they are called pseudo-rotaxanes, when the bulkier end
beads are not present.\cite{Sevick_10,Araki_07} Novel materials called
topological gels, have been produced with melts of polyrotaxanes,
by cross-linking two $\alpha$-cyclodextrin molecules belonging to
different rotaxanes.\cite{Noda_14}These eight-shaped links are then
movable through the backbone of the polymers, unlike the chemical
gels, which have fixed cross-links. The topological gels refer usually
to polymer melts in a solvent, but the polyrotaxanes aggregates with
movable cross links can also be produced with a dry polymeric matrix,
receiving the name of slide-ring materials\cite{Noda_14,Ito_10,Kato_14}. 

\begin{figure}[h]
\begin{centering}
\includegraphics[clip,width=0.98\columnwidth]{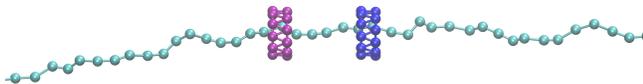}
\par\end{centering}
\caption{\label{fig:chain_ring_sketch}Snapshot of the system close to the
rings, as given by the simulations. The rings are fixed at distance
$d$. The polymer chain fluctuates at thermal equilibrium in the canonical
ensemble. A typical chain has $N=1024$ beads and the chain ends are
connected through periodic boundary conditions. The links show bond
connections for rings and chain given by the FENE model (see Sec.
\ref{sec:Simulation-techniques}). }
\end{figure}

In this work we study the properties of a fluctuating chain under
tension with and without threaded rings fixed at given positions ,
which act as physical constraints of the fluctuation of the chains.
The physical system we deal with is shown in Fig. \ref{fig:chain_ring_sketch},
with a typical configuration of the chain-ring system. For the polymer
chain without rings, we find a transversal fluctuation spectrum compatible
with $\sim1/q^{2}$ dependence at high stretchings, and a deviation
from it at shorter chain stretchings. We show how this spectrum is
modified by the presence of two rings fixed in space, and the appearence
of an attractive fluctuation-induced force between the rings. The
details of the model and simulation techniques are explained in Section
\ref{sec:Simulation-techniques} and we present the results for the
fluctuation properties of the chain as a function of chain extensions
and temperatures in Section \ref{subsec:chains_without_rings}. We
devote section \ref{subsec:chain-with-rings} to the characterization
of the fluctuation-induced forces between the rings as a function
of chain extension and temperature. We also compare the dependence of 
the force with the rings' distance with models of different fields, such 
as classical electromagnetism and liquid-liquid interfaces.
In Section
\ref{sec:Conclusions}, we provide a final discussion and conclusions.

\section{Model and Simulation techniques\label{sec:Simulation-techniques}}

We use the widely known and studied Kremer-Grest model\cite{Grest_86,Kremer_90}
for the polymer chain under tension. The interaction between neighboring
beads along the polymer chain is modeled by a finitely extensible
non-linear elastic (FENE) potential:

\begin{equation}
U_{{\rm FENE}}=\begin{cases}
-\frac{1}{2}k\,\,R_{0}^{2}\ln\left[1-\left(\frac{r_{ij}}{R_{0}}\right)^{2}\right] & \mbox{for}\,r_{ij}\leq R_{0}\\
\infty & \mbox{for}\,r_{ij}>R_{0}
\end{cases},\label{eq:FENE potential}
\end{equation}
 where the maximum allowed bond length is $R_{0}=1.5\sigma$, the
spring constant is $k=30\varepsilon/\sigma^{2}$, and $r_{ij}=|{\mathbf{r}_{i}}-{\mathbf{r}_{j}}|$
denotes the distance between neighboring monomers. Excluded volume
interactions at short distances and van-der-Waals attractions between
beads are described by a truncated and shifted Lennard-Jones (LJ)
potential:

\begin{equation}
U(r)=U_{{\rm LJ}}(r)-U_{{\rm LJ}}(r_{{\rm c}})\,,
\end{equation}
 with

\begin{equation}
U_{{\rm LJ}}(r)=4\varepsilon\left[\left(\frac{\sigma}{r}\right)^{12}-\left(\frac{\sigma}{r}\right)^{6}\right]\,,
\end{equation}
 where the LJ parameters, $\varepsilon$ and $\sigma$, define the
units of energy and length, respectively. Temperature is given in
units of $\varepsilon/k_{B}$, with $k_{B}$ being the Boltzmann constant.
$U_{{\rm LJ}}(r_{{\rm c}})$ is the LJ potential evaluated at the
cut-off radius. We used standard values for the LJ parameters and
mass: $\sigma=1$, $\varepsilon=1$ and $m=1$. The interaction cut-off
is located at the minimum of the LJ potential, $r_{{\rm c}}=2^{\frac{1}{6}}\sigma$,
which gives effectively a fully repulsive potential and is typical
of good solvent conditions when studying polymer melts\cite{Pastorino_06,Pastorino_09}.
This model has been applied to a variety of thermodynamic conditions,
chain lengths, and physical regimes such as glasses, melts, dilute
solutions, etc.\cite{grest_review_brush,review_baschnagel_varnik,Kroeger_04}.
It has been also used for the study of single polymer chains under
tension, in a similar physical situation as studied here, with focus
in dynamical and relaxation properties.\cite{Febbo_08} The polymer
chain was maintained under tension by connecting the beads 1 and $N$
through a FENE potential. Periodic boundary conditions were applied
by computing the force on each of these beads with the periodic image
of the other one. In this way, the length of the chain is set with
the box dimension $L_{x}$ in the $\hat{x}$ direction. In addition,
the center of mass of the chain was kept fixed at the center of the
MD box. 

The ring molecules were generated at fixed positions within the boundaries
of the simulation box, and the positions of their beads were not allowed
to evolve in time. They were modeled by groups of 11 beads, arranged
in two parallel circles. This number was chosen because we found it
to be the minimal number of beads at which the rings did not unthread
from the chain, for different chain stretchings and temperatures.
We chose the beads of the rings to be the same size than the beads
of the chain for simplicity. We do not aim at developing a detailed
model of anular molecule. The ring molecules are meant to provide
a consistent physical constraint for the transverse fluctuations of
the polymer chain. Beads in the rings are connected by springs (see
Fig. \ref{fig:chain_ring_sketch}) with the same FENE interactions
that we used for the connectivity of the chains. In the model of the
ring molecules it was necessary to use two interlocked groups of beads,
because setting only one is prone to unthreading from the chain at
high stretchings or temperatures (see Fig. \ref{fig:chain_ring_sketch}).
The excluded volume of beads was also described by a LJ potential,
also with a cut-off of $r_{{\rm c}}=2^{\frac{1}{6}}\sigma$ which
keeps only the repulsive part of the LJ potential. The interaction
of the rings with the chain is therefore purely repulsive, as well
as the interaction among beads of the rings. This rules out any direct
attractive interaction between rings, which is important to isolate
the effective interaction arising from the Casimir-like forces. 

We used a Langevin thermostat to study the system at constant temperature.
Dissipative and stochastic forces are added to the conservative forces,
already present in the standard molecular dynamics equations of motions.
The dissipative force on particle $i$ is given by ${\mathbf{F}_{i}}^{{\rm D}}=-\gamma\mathbf{v}_{i}$,
where $\gamma$ is the friction coefficient and $\mathbf{v}_{i}$
the particle velocity. The random force, $\mathbf{F}_{i}^{{\rm R}}$,
has zero mean value and its variance satisfies\cite{Hunenberger_05,Pastorino_07}

\begin{equation}
\langle F_{i\mu}^{{\rm R}}(t)F_{j\nu}^{{\rm R}}(t^{\prime})\rangle=2\gamma Tk_{B}\delta_{ij}\delta_{\mu\nu}\delta(t-t^{\prime})\,,\label{eq:mean-square-random-lgv}
\end{equation}
 where the indices $i$ and $j$ label particles, $\mu$ and $\nu$
Cartesian components, and $T$ is the temperature at which the system
is simulated.

After a thermalization stage of $1\times10^{6}$ MD steps with a time
step of $dt=1\times10^{-5}\tau$, typical simulations were performed
with trajectories of $1\times10^{7}$ MD steps each, with a time step
of $dt=2\times10^{-3}\tau$. The time unit in LJ parameters being
$\tau=\sigma(m/\varepsilon)^{1/2}$. We took averages of physical
quantities each 1000 time steps. The friction constant was set at
$\gamma=0.5\varepsilon\tau/\sigma^{2}$ for all the simulations, except
for the initial thermalization stage in which the friction constant
was set to $\gamma=50.0\varepsilon\tau/\sigma^{2}$. The typical chain
was composed of $N=1024$ beads, but we also studied shorter and longer
chains, when needed. These cases will be mentioned explicitly in the
text. This choice allowed us to study ring distances in the range
$3-60\sigma$, to obtain a fluctuation-induced force values with high
enough signal-to-noise ratio. 

\section{Results}

\subsection{Properties of the polymer chain under tension \label{subsec:chains_without_rings}}

\begin{figure}[t]
\begin{centering}
(a) $L^{*}=0.40$
\par\end{centering}
\begin{centering}
\includegraphics[clip,width=0.98\columnwidth]{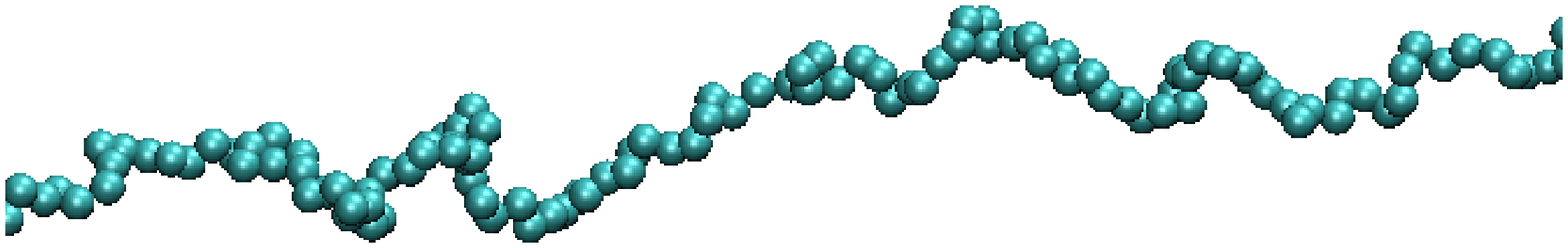}
\par\end{centering}
\begin{centering}
(b)$L^{*}=0.57$ \includegraphics[clip,width=0.98\columnwidth]{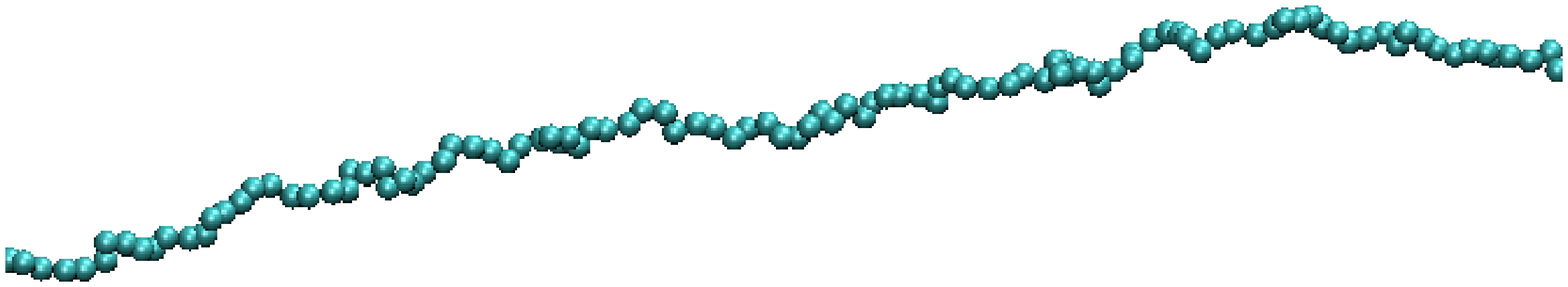}
\par\end{centering}
\begin{centering}
(c)$L^{*}=0.64$ \includegraphics[clip,width=0.98\columnwidth]{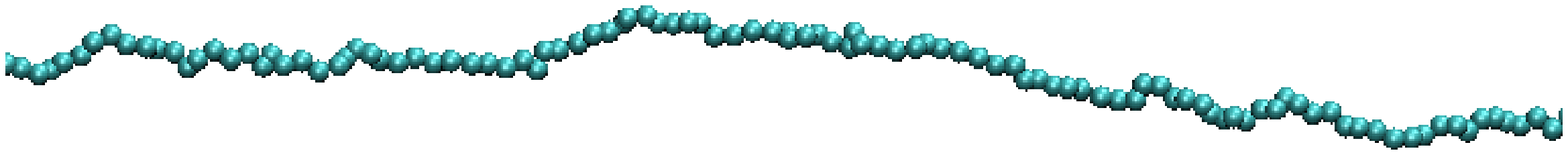}
\par\end{centering}
\begin{centering}
(d)$L^{*}=0.70$
\par\end{centering}
\begin{centering}
\includegraphics[clip,width=0.98\columnwidth]{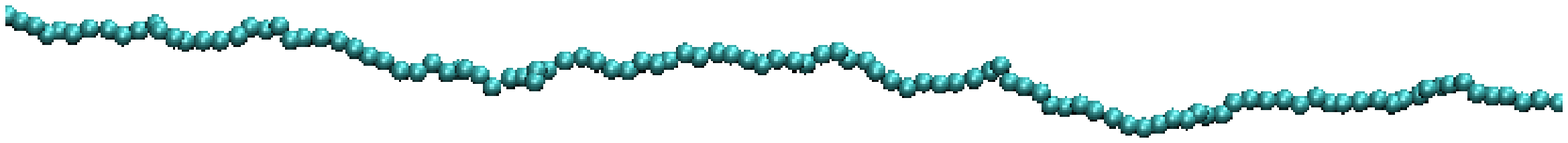}
\par\end{centering}
\begin{centering}
(e)$L^{*}=0.80$ \includegraphics[clip,width=0.98\columnwidth]{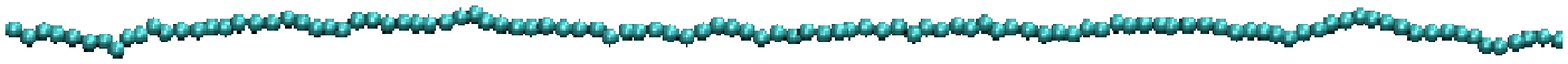}
\par\end{centering}
\begin{centering}
(f)$L^{*}=0.90$ 
\par\end{centering}
\begin{centering}
\includegraphics[clip,width=0.98\columnwidth]{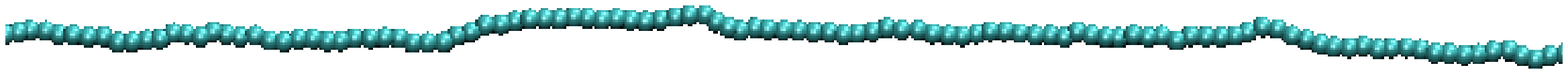}
\par\end{centering}
\centering{}\caption{\foreignlanguage{spanish}{\label{fig:Snapshots-chain-extensions}\foreignlanguage{american}{Snapshots
of a fragment of the polymer chain at different extensions and temperature
 $T=33.6\varepsilon/k_{B}$. $L^{*}$accounts for the length as a
fraction of the maximum possible chain extension in the FENE model
$L_{max}=1.5\sigma N$. }}}
\end{figure}

We analyze firstly the structural and fluctuation properties of the
chain under extensional force, without rings. We worked with simulations
at constant length, giving rise to a mean constant stretching force.
Fig. \ref{fig:Snapshots-chain-extensions} shows typical configurations
of the chains for different extensions. The chain length is given
in units of the maximum extension of the FENE model chain \textbf{$L_{{\rm max}}=1.5N\sigma$}.
For the smaller extensions the chain starts to show blog formation.
There are local regions of beads arranged in configurations not very
different from that of isolated chains (Fig. \ref{fig:Snapshots-chain-extensions},
$L^{*}=0.4$). For larger extensions the chain is very stretched,
with little freedom for transverse relative displacement among consecutive
beads (Fig. \ref{fig:Snapshots-chain-extensions}, $L^{*}=0.70$).
In this regime the internal energy of the chain is very high, and
dominant as compared to configurational entropy. 

\begin{figure}[t]
\begin{centering}
\includegraphics[clip,width=0.98\columnwidth]{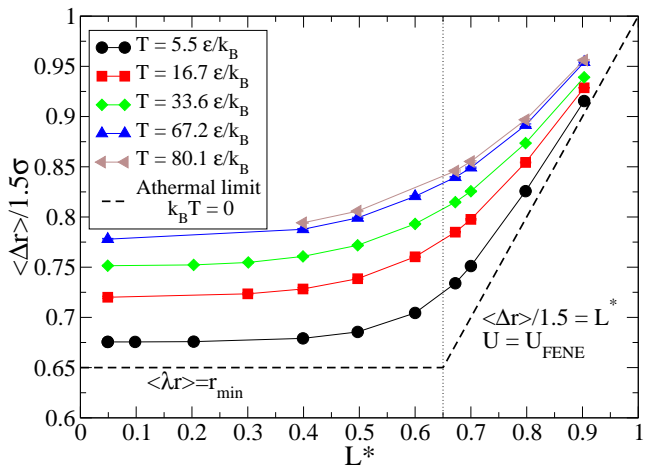} 
\par\end{centering}
\caption{Mean bond length as a function of chain stretching for different temperatures
in units of maximum FENE length. The dashed line represents beads
distance in the limit of $k_{B}T=0$ and no excitations in the chain
(athermal limit). The vertical dotted line indicates the extension
at which the bond length increases its rate of change with chain stretching
\label{fig:Mean-bond-length}}
\end{figure}

\begin{figure}[t]
\begin{centering}
\includegraphics[clip,width=0.98\columnwidth]{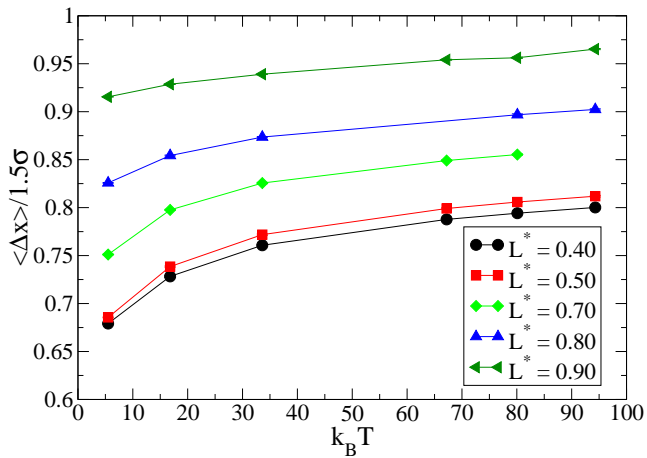}
\par\end{centering}
\caption{Mean distance among beads as a function of temperature for different
chain extensions. A saturation at higher temperatures is observed
in accordance with the fast increase of FENE bond energy.\label{fig:Variaci=0000F3n-distancia-vs_Temp}}
\end{figure}

This behavior can be quantified with the mean bond length in units
of the maximum allowed bond length for the FENE model ($R_{{\rm max}}=1.5\sigma)$.
This is shown in Fig. \ref{fig:Mean-bond-length} as function of chain
stretching, for different temperatures. For stretchings in the range
$L^{*}<0.65$, the mean bond is marginally dependent on chain length,
and increases towards the limit $L*\simeq0.65$. In this regime the
blobs are dominant and bond length values are dominated by the thermal
energy of the beads and the excluded volume which, given by the Lennard-Jones
potential. In the case $L^{*}\gtrsim0.65$, the bonds increase more
pronouncedly with chain stretching. In this limit each bond is permanently
stretched with respect to the equilibrium bond length and the excluded
volume is not important for the mean bond. For very high $L^{*}$,
the bonds converge to the maximum value for all the temperatures.
The case $T=0$ (athermal) is shown for comparison with a dashed line.
This is the limiting case, in which there is only potential energy
in the chain. It follows the tendency of the thermal case, indicating
also the two distinctive behaviors. The value $r_{{\rm min}}\sim0.65$
corresponds to a bond distance $d=0.975\sigma$, where the Lennard-Jones
and bond forces are equal. Fig. \ref{fig:Variaci=0000F3n-distancia-vs_Temp},
shows the bond distance as a function of temperature for different
chain extensions. The mean bond is more dependent on temperature at
shorter chain extensions and lower temperatures. The FENE contribution
of the bond energy is, of course, non-linear, and therefore the temperature
has a progressively minor effect with increasing temperatures. 

We also calculate the spectra of transverse fluctuations of the chain
(i.e. those perpendicular to the stretching direction $\hat{x}$ ),
to characterize the collective vibrations in the limit of high stretching.
We point out that even the lowest stretching case ($L^{*}=0.20$)
is still very high, as compared with a free polymer chain. We define
a discrete function $h(x_{i})=h_{bin}(x_{i})-h_{0}$, which accounts
for the transverse fluctuations. $h_{bin}$ is obtained by dividing
the space along the chain in bins of width $\Delta x=2\sigma$ and
then computing the mean position of the beads that belong to each
bin. This binning procedure was used because for short chain lengths
the beads are grouped in blobs and taking directly their positions
would give a multivalued function at some $x$ points. This type of
discretization is usual in analysis of interface fluctuations\cite{Vink_05,Pastorino_09}.
$h_{0}$ is the horizontal along $x$ direction, in which would lie
the stretched chain at zero kinetic energy. For each time step the
Fourier amplitudes are calculated and averaged over the chain configurations
obtained in the simulations. The fluctuation espectra, given by the
square amplitude of the Fourier modes, are shown in Fig. 5, for some
selected chain lengths. The inset shows the power spectra in logarithmic
scale. The overall magnitude of fluctuations is, as expected, reduced
with chain stretching. The wave amplitudes also decrease with $q$
number or, equivalently, increase with wavelength $\lambda$.

\begin{figure}[h]
\begin{centering}
\includegraphics[clip,width=0.98\columnwidth]{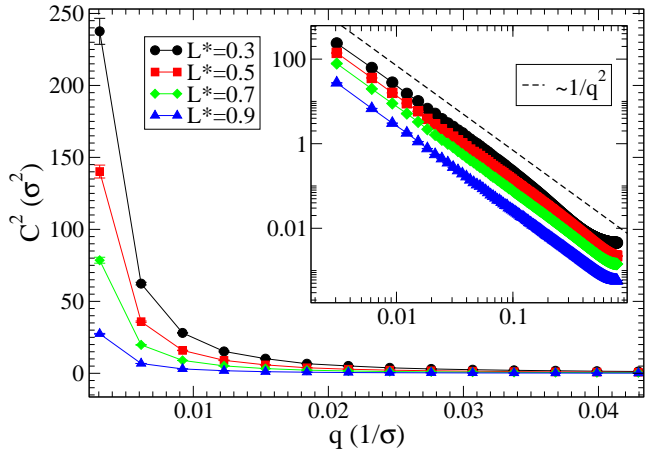} 
\par\end{centering}
\caption{\foreignlanguage{spanish}{\label{fig:Fourier-spectrum-of-the-chain}\foreignlanguage{american}{Fourier
spectrum of the polymer chain at different lengths for $T=33.6\varepsilon/k_{B}$.
The inset shows the same data in log-log plot. }\protect \\
}}
\end{figure}

In the limit of very high stretching, the harmonic approximation should
fulfill, due to small amplitude oscillations of the beads, which are
effectively trapped in very stiff potential wells. This can be thought
as an effective Hamiltonian with quadratic degrees of freedom in coordinates
and momenta. The chain is in a heat bath at constant temperature and
therefore, equipartition theorem holds, giving a contribution of $\frac{1}{2}k_{B}T$
for each degree of freedom to the potential and kinetic energies.
Each normal mode of the chain has the same mean potential energy of
$\frac{1}{2}k_{B}T$ and, by using the relation between potential
energy of a mode and amplitude\cite{Pain_05}, the harmonic model
gives rise to a dependence of the squared mode amplitude $C^{2}\sim1/q^{2}$.
The same conclusion can be reached in the realm of soft matter and
interfaces for a capillary wave hamiltonian. The fluctuating stretched
chain projected in a plane, can be thought as a unidimensional interface
between two immiscible liquids or a liquid-gas interface and the modes
of the chain as the capillary waves of the interface. The energy cost
of a non-flat surface in comparison to the flat case (of minimum area)
can be written as an effective Hamiltonian of surface fluctuations.
This so-called capillary wave Hamiltonian, describes the energy cost
of surface undulations of thermal origin in terms of a function $h(x,y)$,
which accounts for the local position of the interface. Expressing
this Hamiltonian in Fourier space, leads to a quadratic form in the
wave vectors $q$ of independent harmonic oscillators. The application
of the equipartition theorem, leads to the dependence $C(q)\text{\ensuremath{\sim}}k_{B}T/q^{2}$
for the Fourier amplitudes of the Fourier modes of the surface.\cite{Vink_05,Safran_03}
This analogy is very interesting, because simulations can test in
which range of chain extensions is valid and allows the study of interface
fluctuations by carefully simulating a stretched chain. The chain
is, of course, much less demanding of computing power. The dashed line  
  in the inset of Fig. \ref{fig:Fourier-spectrum-of-the-chain}, shows a curve $\sim1/q^{2}$ 
for reference. For the higher extensions ($L^{*}=0.90$ and $L^{*}=0.70$)
the chain fluctuations are very close to the harmonic model. The curves
are parallel to $1/q^{2}$ for the whole range of $q$. Some differences
show up for the smaller extension $L^{*}=0.40$, which increase appreciably
for the lowest stretching $L^{*}=0.20$. This is specially true for
the high-$q$ part of the spectra, i.e. for the shorter-wavelength
modes. Interestingly this is in line with the idea of the capillary
wave Hamiltonian (also the Helfrich Hamiltonian, for membrane bilayers).
These models are suitable for long-wavelength fluctuations\cite{Safran_03,Vink_05}.
We recall that for the shorter chain lengths, the Lennard-Jones interaction,
i.e. the excluded volume, has a role in the local dynamics of the
beads.

\begin{figure}[t]
\begin{centering}
\includegraphics[clip,width=0.98\columnwidth]{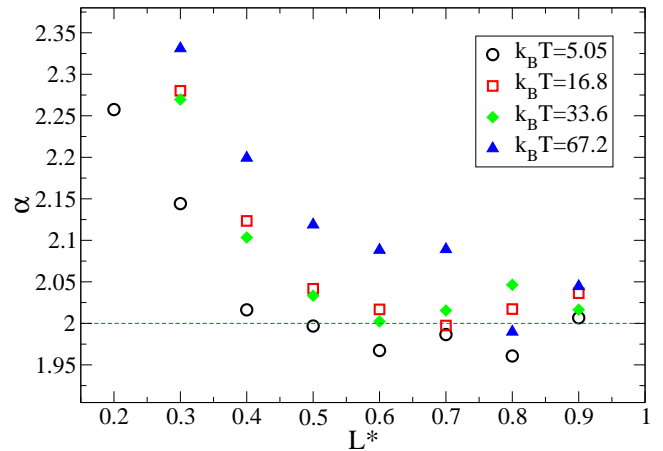} 
\par\end{centering}
\caption{Exponent of the fitted decay of the fluctuation spectrum as function
of chain length for different temperatures . From the capillary-wave
hamiltonian a coefficient $\alpha=2$ is expected (indicated with
dashed line). The exponent gets closer to $\alpha=2$ from $L^*=0.5$ toward 
higher stretching values. \label{fig:Exponent-capillary-wave}}
\end{figure}

Fig. \ref{fig:Exponent-capillary-wave} show the decaying exponent
of the spectra for different chain stretchings and temperatures. We
fitted the function $f(q)=A_{0}/q^{\alpha}$ in the log-form, where
$A_{0}$ and $\alpha$ are the fitting parameters. A convergence towards
$\alpha=2$ is observed upon increase of chain stretching. For the
range of shorter chain stretchings ($L^{*}<0.5$) the exponent increases.
This means a deeper decay of fluctuations for smaller wavelengths.
For these lower stretchings, the polymer chains are more intertwined
and the vibrational behavior comes dominantly from groups of beads,
more than from independent beads themselves. This could be thought
as an effectively shorter chain (with fewer degrees of freedom and,
therefore modes), which for the same $q$ range of the complete chain,
it will have a faster decay. 

\subsection{Effect of the rings on the chain and fluctuation-induced forces \label{subsec:chain-with-rings}}

\begin{figure}[t]
\begin{centering}
\includegraphics[clip,width=0.98\columnwidth]{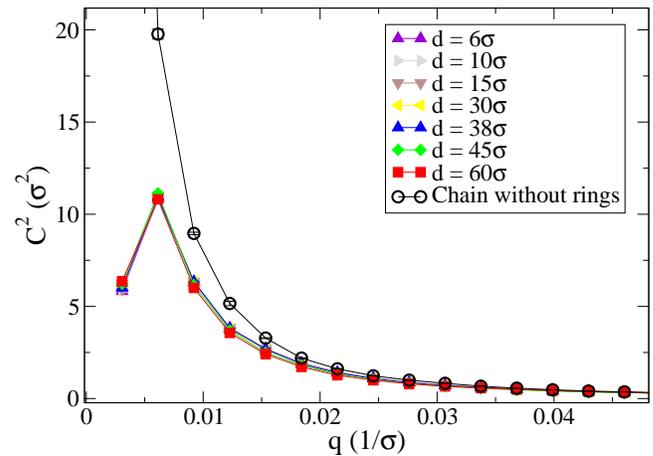}
\par\end{centering}
\caption{Fourier spectrum for the chain without rings (open circles) and with
the rings located at different distances (filled symbols). All the
cases correspond to $T=33.6\varepsilon/k_{B}$ and $L^{*}=0.70$.
The effect of the rings is only noticeable at small $q$ (long wavelengths)
and it becomes unnoticeable at high  $q$ values. \label{fig:spectra_amp_w_rings}}
\end{figure}

We analyze here the effect of the fixed rings in the natural fluctuations
of the chain, to continue further to the fluctuation-induced or Casimir-like
forces between rings. Fig. \ref{fig:spectra_amp_w_rings} shows the
fluctuation spectra of the chain with the rings fixed at different
distances. The decay range is similar for the chain with and without
rings, but the modes whose wavelengths are higher than the distance
between rings ($\lambda_{{\rm mode}}>d$) are significantly reduced.
The power spectrum of the chain without rings is shown also for comparison. 

\begin{figure}[t]
\begin{centering}
\includegraphics[clip,width=0.98\columnwidth]{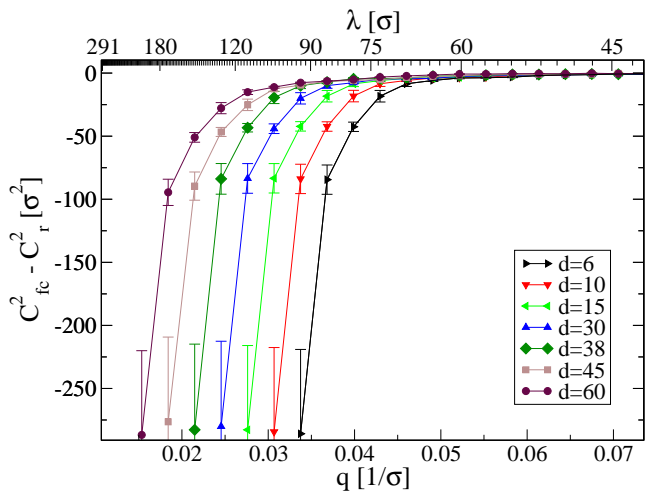}
\par\end{centering}
\caption{\label{fig:resta_espectro}Difference of Fourier coefficient normalized
sum for the case of the chain with rings ($C_{r}^{2}$) and the chains
without rings ($C_{fc}^{2}$). The legends show the distance between
rings at which the spectra were calculated. The vertical line indicates
the $q$ value ($\lambda\sim67\sigma)$ below which the presence of
rings begins to influence the amplitude of the chain modes. The cut-off
of modes at $q\lesssim2\pi/d$ produced by the presence of the rings
is clearly appreciated. }
\end{figure}

Fig. \ref{fig:resta_espectro} shows the difference of the power spectrum
with and without rings for different ring distances. The effect of
the rings is clearly observed for the modes which are expected to
be heavily hindered by the rings. The higher effect is observed for
smaller distances in which modes of lower wavelengths are reduced,
starting at $\lambda_{min}=d$. It is also observed that increasing
the ring distance, changes the cut-off wave number from which the
spectrum is significantly reduced, as compared to the stretched chain
without rings. 

\begin{figure}[t]
\begin{centering}
\begin{tabular}{c}
(a)\includegraphics[clip,width=0.98\columnwidth]{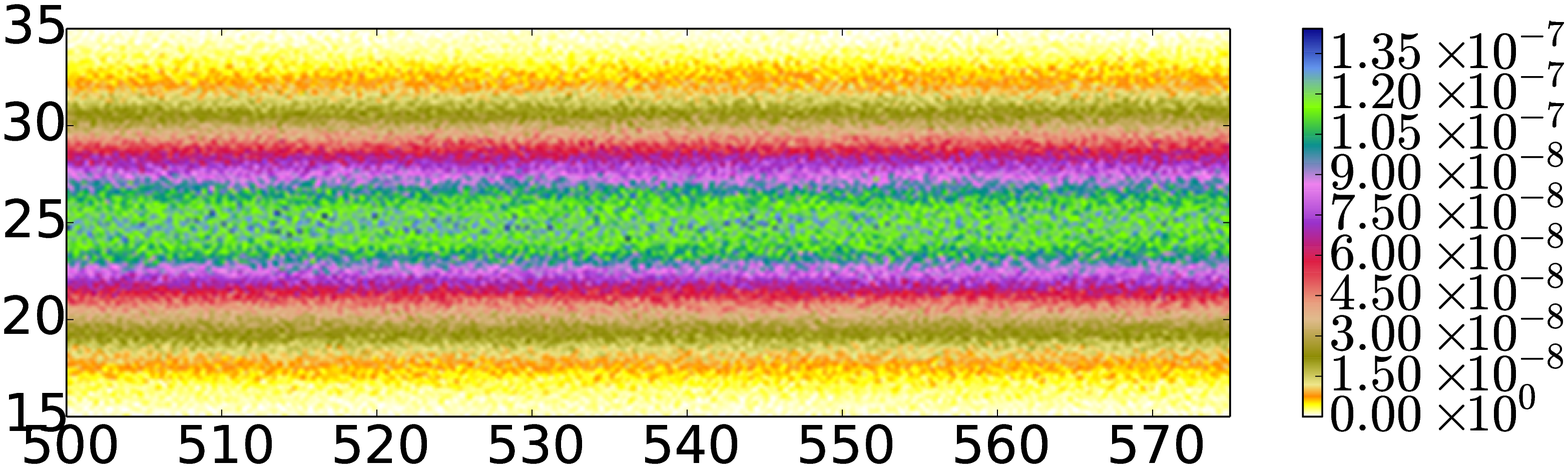}\tabularnewline
(b)\includegraphics[clip,width=0.98\columnwidth]{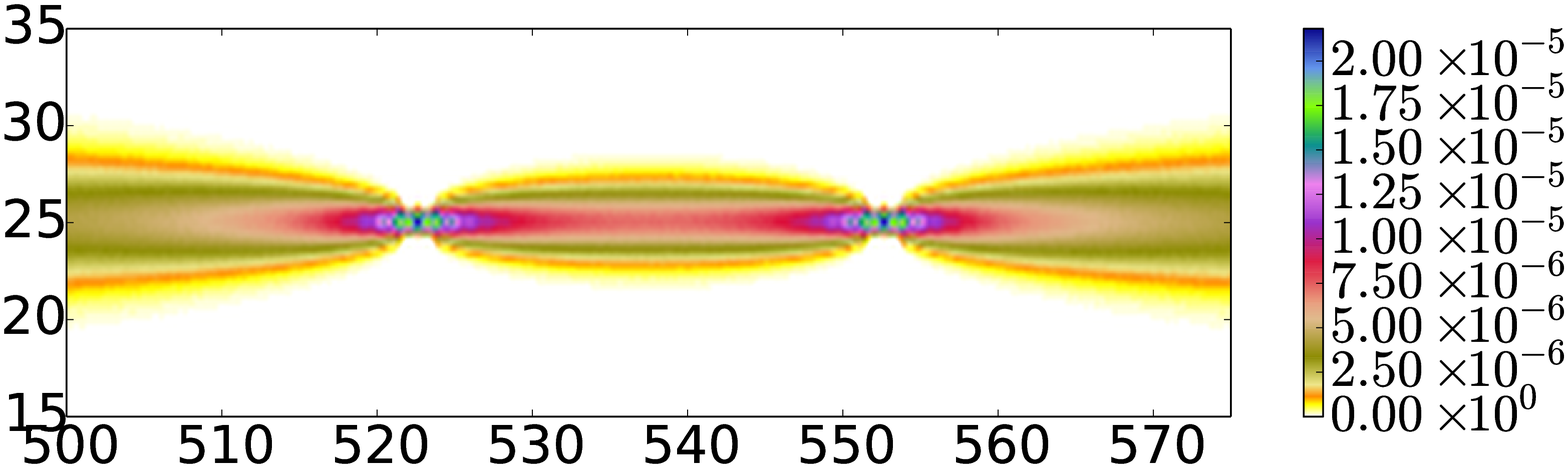}\tabularnewline
(c)\includegraphics[clip,width=0.98\columnwidth]{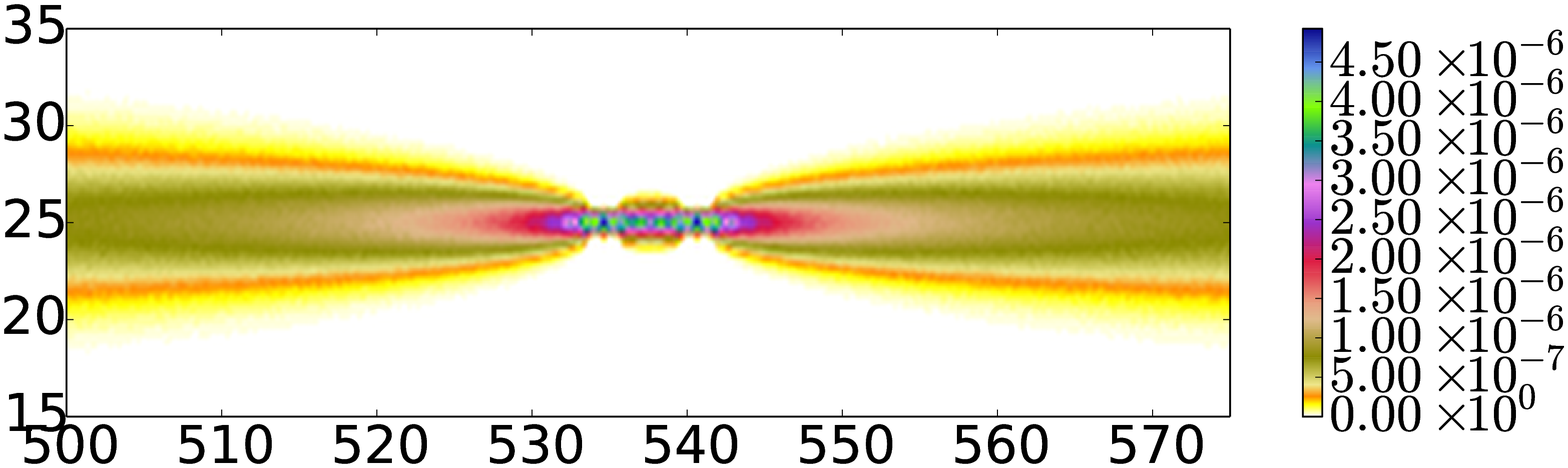}\tabularnewline
\end{tabular}
\par\end{centering}
\caption{Color plot of the number density for the central zone of the chain.
Panel (a) presents the chain without rings. The density with rings
at distance $d=30\sigma$ is presented in panel (b) and with rings
at $d=6\sigma$ are presented in panel (c)\label{fig:Density-plot}}
\end{figure}

In Fig. \ref{fig:Density-plot} the mean number density of the chain
is presented. Panel (a) shows the chain without rings, while panels
(b) and (c) present the changes for rings located at distances $30\sigma$
and $6\sigma$ respectively. We obtained the histograms from a square
binning in 2D with a bin lateral size of $\Delta r=0.25\sigma$ and
a Bessel smoothing function was used for the color plots. The spatial
zone at which the beads have access is significantly reduced by the
presence of the rings. This is in line with the suppression of modes
of higher amplitude in the presence of rings. For the zone between
rings, the bead distribution is similar to the outer zone for high
ring distances (Panel B, $d=30\sigma$), but with reduced amplitudes.
For rings very close (Panel C, $d=6\sigma$), the inner zone presents
beads only very close to the rings. The outer zone however, presents
monomers in a slightly wider zone as compared to the case of rings
located at higher distances (see Panel B, in Fig. \ref{fig:Density-plot}),
compatible with the fact that the outer zone of the chain can have
modes of very high wavelength and amplitude (see Sec. \ref{subsec:chains_without_rings})
when the rings are very close to each other. 

\selectlanguage{spanish}%
\begin{center}
\par\end{center}

\begin{figure}[t]
\begin{centering}
\includegraphics[clip,width=0.98\columnwidth]{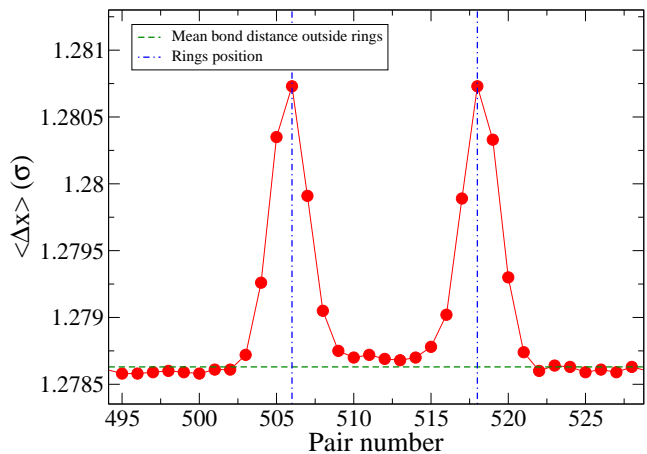}
\par\end{centering}
\caption{\foreignlanguage{spanish}{\label{fig:bond_length_pair}\foreignlanguage{american}{Mean bond
lengths $\langle\Delta x\rangle$ as function of bond number for different
polymer lengths. The rings are located at $d=15\sigma$ with a chain
stretching of $L^{*}=0.8$ and temperature $T=15\varepsilon/k_{B}$.
The bond number is labeled such that bond number $b_{i}$ is the bond
distance between beads $i+1$ and $i$. Vertical dot-dashed lines
indicate the approximate position of the rings. The error bars have
almost the same size than the symbols. }}}
\end{figure}

\selectlanguage{american}%
In addition to the fluctuation properties of the chain, we found other
interesting aspect that changes in the presence of fixed ring molecules.
We studied the mean bond length for all the chain bonds, considering
separately the bonds lying between the rings and those in the outer
region. Fig. \ref{fig:bond_length_pair} depicts the mean bond lengths
obtained from the mean bond value over neighboring beads in the chain.
The pair number is defined such that pair number $i$, indicates the
bond between beads $i+1$ and $i$. Fig. \ref{fig:bond_length_pair}
provides an example of what we have observed for all the cases. First,
there is a significant stretching of the bonds which are directly
exposed to the fixed rings. This can be expected: as the chain fluctuates
the bond will scatter against the beads of the rings, producing a
high stretching of these bonds. There is also a much more subtle and
interesting effect, that we noticed for different stretchings and
temperatures. Namely, the bond between rings are slightly stretched
as average. This can be observed in the center points of Fig. \ref{fig:bond_length_pair}
which are systematically above the dashed line which indicates the
mean bond value for the bonds in the outer region of the chains (those
in the regions of bond numbers 1-503 and 522-1023). The chain is more
stretched between rings. From a mechanical viewpoint, we attribute
this to action-reaction principle. As it will be shown next, the chain
is effectively producing a mean force between rings and there should
be a force equal in magnitude that the rings exert on the chain, giving
rise to this slight increase of the bond length in the zone between
rings. This interesting characteristic could be explored further,
but this is outside the scope of the present work. 

\begin{figure}[h]
\begin{centering}
\includegraphics[clip,width=0.98\columnwidth]{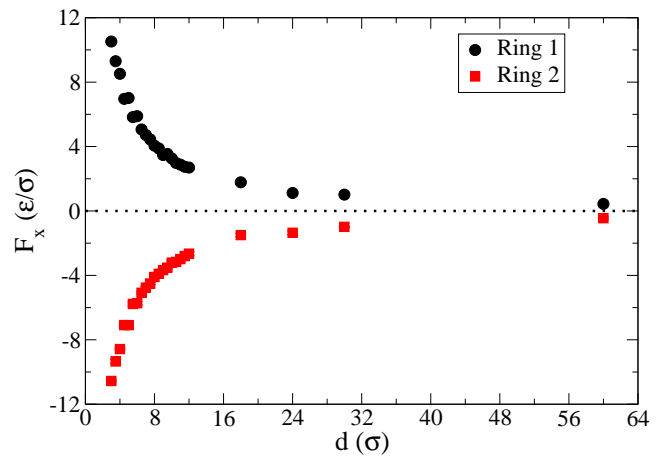} 
\par\end{centering}
\caption{\label{fig:f_ring_vs_d_typical}Total mean force on rings $\langle f_{ring}^{(x)}\rangle$
in the chain direction as a function of ring distance. Each annular
molecule is composed of two rings formed by 11 LJ particles located
on a circle of radius $1.5\sigma$. The sample was set at $T=33.6\varepsilon/k_{B}$
and $L^{*}=0.70$. These are the fluctuation-induced or Casimir-like
forces from the chain on the anular molecules, due to the restrictions
that they impose in the natural fluctuations of the chain. For a given
distance the rings hinder the fluctuations of the chain with modes
of wavelength $\lambda>d$. }
\end{figure}

The nature and characteristics of effective forces between rings is
one of the main results and motivations of our study. We compute the
mean force on each bead of the rings, averaged over each MD step.
This quantity is the mean force on the ring due to the interaction
with the chain. We recall that ring-ring interaction is neglected
because we used a cut-off for the ring-ring interaction of $R_{c}=1.12\sigma$,
among beads of same and different rings. We kept only the repulsive
part of the Lennard-Jones potential, which means that only the excluded
volume is considered. As rings were fixed at distances $d\geq3\sigma$,
direct interaction between rings are disregarded by construction.
We show in Fig. \ref{fig:f_ring_vs_d_typical} one of the main results
of this work, namely the existence of an effective force between rings
that can only arise through interactions mediated by the chain. Furthermore,
the rings are efficient to hinder transverse fluctuations of the chains,
but not longitudinal waves, since the effective ring diameter is bigger
than the bead diameter of the chain ($1\sigma$). We conclude that
this effective interaction is the expected fluctuation-induced force
due to the disturbance of the natural fluctuations of the stretched
unconstrained chain by the presence of the rings. They have a range
clearly larger than ring size ($\sim6\sigma$). Another qualitatively
important aspect is that the interaction is attractive for all the
studied cases. We point out also that the nature of the system disregards
other type of effective interactions, as for example, depletion interactions
which are usually present together with fluctuation-induced forces
in many systems\cite{israelachvili_11}. From a thermodynamic viewpoint,
the system minimizes free energy with the rings as close as possible
to each other. This allows for modes with longer $\lambda$ in the
outer region. However, as we have the chain in a thermal bath (canonical
ensemble) the mean internal energy of the chain is the same for all
the cases. Therefore the Casimir-like force arises in particular from
the maximization of entropy, which is obtained when the rings are
together. Entropy is maximized, when the maximum possible number of
modes are active and specially those of higher amplitude (high $\lambda$,
see section \ref{subsec:chains_without_rings}). As we showed, these
modes are hindered when the rings are placed at higher distances. 

\begin{figure}[t]
\begin{centering}
~\includegraphics[clip,width=0.98\columnwidth]{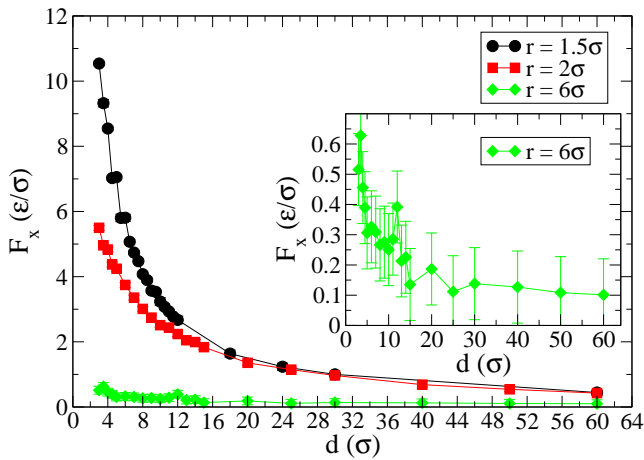}
\par\end{centering}
\caption{\label{fig:Mean-force-diff-radius}Mean force between rings versus
distance for different ring radius $r$. The parameters were set $L*=0.70$
and $T=33.6\varepsilon/k_{B}$. The Inset shows a magnification of
the case $r=6\sigma$, for which the minimal restrictions imposed
by this big ring radius reduces significantly the fluctuation-induced
force. }
\end{figure}

The ring molecules, are of course physical constraints and not mathematical
nodes imposed on the chain. It is helpful to analyze the fluctuation-induced
force as a function of ring radius. This is done in Fig. \ref{fig:Mean-force-diff-radius},
where we plot force versus ring distance for different ring radius.
The force is present in all the cases, but greatly reduced for larger
ring radii. The inset shows a zoom of the higher studied radius ($r_{{\rm ring}}=6\sigma$),
which is barely noticeable in the scale of the main graph. As the
ring radius increases, the rings are unable to hinder the fluctuations
of middle to small wavelength, which are progressively of smaller
amplitude as $\lambda$ decreases. The ring radius modifies significantly
the strength of the fluctuation-induced force. 

\begin{figure}[t]
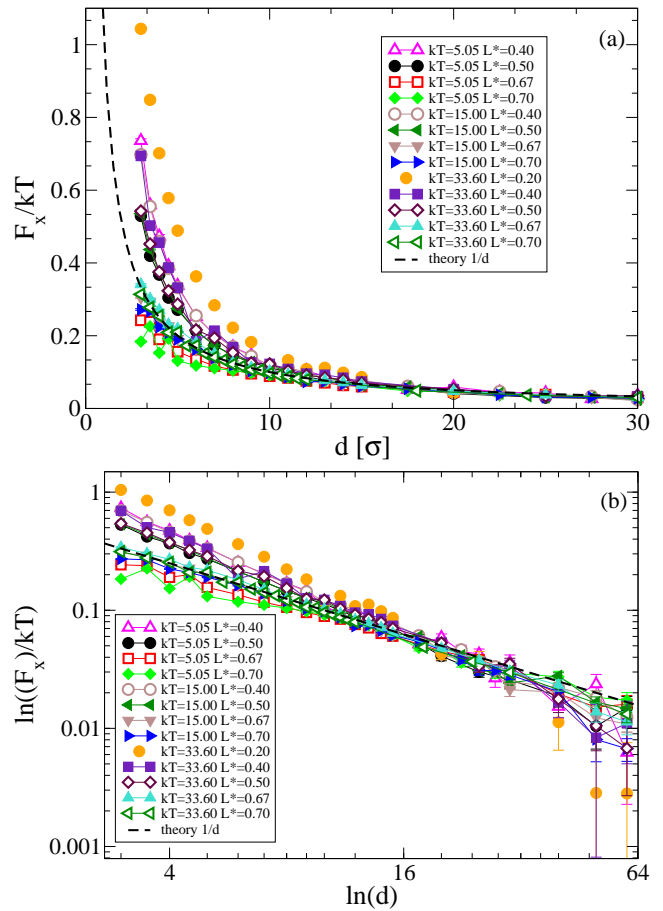

\begin{centering}
\begin{tabular}{c}
\includegraphics[clip,width=0.98\columnwidth]{./Div_allT_allL}\tabularnewline
\includegraphics[clip,width=0.98\columnwidth]{./log_log_Div_allT_allL}\tabularnewline
\end{tabular}
\par\end{centering}
\selectlanguage{american}%
\caption{\label{fig:f_vs_d_different-stretching}Upper panel: Mean force versus
ring distance for different chain stretchings and temperatures. The
dashed line represents the harmonic model for the force $f\sim1/d$.
It indicates only the power law decay. Lower panel: log-log plot for
the same cases. The dashed shows an idealized theoretical model which
is consistent for medium to high stretching (see text).}
\selectlanguage{spanish}%
\end{figure}

We present in Fig. \ref{fig:f_vs_d_different-stretching} the force
strength between rings for different chain stretchings and temperatures
as a function of distance. Panel (a) presents the force intensity
versus distance. The force is long range as compared to bead size.
We resolve non-zero force values up to $d\simeq48\sigma$ for a chain
length of $L=1076\sigma$ and 1024 beads. The force is non\textendash negligible
approximately for distances in a range of $5\%$ of the chain length.
For short ring distances, Fig. \ref{fig:f_vs_d_different-stretching}
shows that the strength of the force is highly influenced by the amplitude
of the chain modes of higher wavelength. As it was showed for the
fluctuations of the chain without rings (see Fig. \ref{fig:Fourier-spectrum-of-the-chain}),
less stretching increases considerably the amplitude of the modes
in the small $q$ range (longer wavelengths). On higher chain stretchings,
for instance $L^{*}=0.8$ in Fig. \ref{fig:f_vs_d_different-stretching}(a),
it appears an oscillation of the force, noticeable at shorter ring
distances. We will analyze this further in the next paragraphs. 

\selectlanguage{spanish}%

\selectlanguage{american}%

\subsubsection*{Comparison with theoretical models}

To compare the force with analytical models, we provide a log-log
plot of the force scaled with the temperature in Fig. \ref{fig:f_vs_d_different-stretching}(b).
Firstly we fitted force between rings as a function of distance as
a power law of the form: $F(d)=A(T)/d^{\alpha}$. A log-log plot should
present a linear dependence, which is approximately the case in Fig.
\ref{fig:f_vs_d_different-stretching}(b). The exponent $\alpha$
changes with chain stretching, growing for smaller $L^{*}$. The dashed
line represents the case $\alpha=1$, which can be obtained analytically
with a harmonic approximation of independent normal modes\cite{Boyer_03c,Boyer_03}.
Boyer discusses the standard Casimir effect from zero point energy
fluctuations of the electromagnetic field and thermal fluctuations
of the classical electromagnetic field in a unified way\cite{Boyer_03c,Boyer_03}.
The latter case is equivalent to the chain in a thermal bath. The
system is described as a one dimensional cavity at a given temperature,
with a partition inside the cavity at position $x$. He calculates
the total force on the partition due to the restriction of modes in
the cavity. Considering harmonic modes, the equipartition limit gives
rise to a contribution to the force of each normal mode of $f_{mode}(\omega,L,T)=\frac{k_{B}T}{L}$,
where $L$ is the length of the cavity. Adding up over all the modes,
the total force on the partition at position $x$ is: 
\[
f(x,L,T)=-\frac{k_{B}T}{2}(\frac{1}{x}-\frac{1}{L-x})
\]

where the two terms indicate attractions to each one of the wall cavities\cite{Boyer_03}.
In our case the boundary condition is periodic and the force between
rings is mapped to the force between the partition and one of the
walls. If the partition is close to one of the walls, the interaction
with the other one is negligible. This would be the limit of two rings
at short distance in comparison with chain length ($d\ll L)$, which
we use in the simulations. The ring distance varies in a range 2-60$\sigma$
in a chain of typical length $L=1076\sigma$. We also add a factor
2, due to two independent fluctuation directions for the chain ($\hat{y}$
and $\hat{z}$), which are locally constrained by the presence of
the rings. We end up therefore with a force dependence given by:
\[
F_{x}(d)\equiv f_{{\rm ring}}(d)=\frac{k_{B}T}{d}\,,
\]
which is plotted in Fig. \ref{fig:f_vs_d_different-stretching} in
dashed line. 

Interestingly the theory agrees pretty well with our results for relatively
high chain stretchings ($L^{*}\gtrsim0.67$) and for all the studied
temperatures. At very high chain stretching $L^{*}>0.7$, there are
force oscillations at short ring distances (considered later on),
but we have again good agreement with the theoretical model at longer
ring distances. It should be noted that the interactions of individual
bonds are non-harmonic elastic terms, given by the FENE potential
(see Eq. \ref{eq:FENE potential}). However, in the limit of high
stretching we consider that the harmonic approximation is reasonable.
We think that this is because each bead of the chain is trapped in
a very stiff potential well, which for moderate temperatures, could
be well approximated by a Taylor expansion of second order. This is
in the same spirit of the small amplitude harmonic approximation of
vibrational modes in a solid at relatively low temperature\cite{Ashcroft_76}. 

In the lower range of chain stretchings ($L^{*}<0.65$) we do not
observe a dependence $\sim1/d$. We note that in this regime not all
the degrees of freedom of the chain are taking part of vibrations.
The low stretching produces local clusters of beads, closer to equilibrium
than to bond stretched states (see Fig. \ref{fig:Snapshots-chain-extensions}(a)).
We attribute to this effective reduction of vibrational modes a steeper
decay of the interaction, as compared to the cases of higher stretchings,
where the harmonic approximation holds. It is also interesting to
note that in this region, the range of the fluctuation-induced force
is reduced, but its absolute value at short ring distance increases.
We assume that this happens because the amplitude of long-wavelength
modes increases significantly for shorter chain stretchings (see Fig.
\ref{fig:Fourier-spectrum-of-the-chain}). It is worth noticing that
the fluctuations of the chain without rings is still closer to $\sim1/q^{2}$,
characteristic of the capillary wave hamiltonian. Lehle et al.\cite{Lehle_06}
obtained analytical results for the dependence of effective forces
in interfaces, induced by capillary wavelike fluctuations. They analyzed
colloids trapped in a liquid-liquid interface with different boundary
conditions colloid-interface and different degrees of freedom of the
colloids. They also treated different shapes as spherical, janus colloids
or disks. The different cases included totally fixed colloids pinned
in the interface, colloids allowed to move vertically and colloids
allowed to move vertically and also to tilt. The case closer to the
system considered here is the fluctuation-induced force between disk
colloids of radius $r_{0}$, pinned in the interface at distance $d$.
For the limiting case $d\gg r_{0}$, they find the following expression
for the fluctuation induced force (see Eq. (7) in Ref. \cite{Lehle_06}):
\begin{equation}
F_{x}(d)=-\frac{k_{B}T}{2}\frac{1}{d\ln(d/r_{0})}+\mathcal{O}(d^{-3})\label{eq:f_ln_colloids}
\end{equation}

\begin{figure}[t]
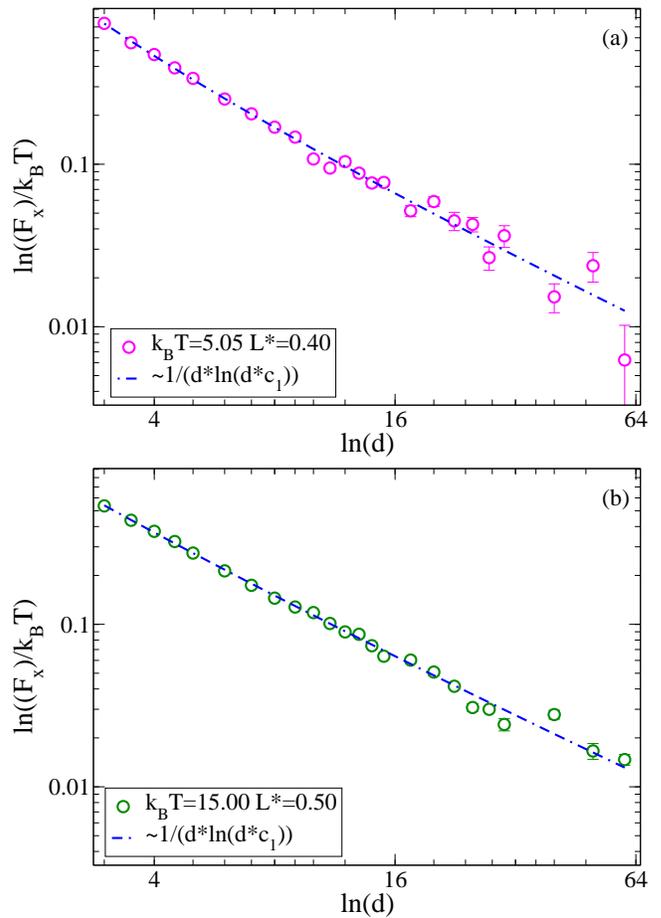

\begin{centering}
\begin{tabular}{c}
\includegraphics[clip,width=0.98\columnwidth]{./set1_log_log_Div_allT_lowL}\tabularnewline
\includegraphics[clip,width=0.98\columnwidth]{./set4_log_log_Div_allT_lowL}\tabularnewline
\end{tabular}
\par\end{centering}
\caption{Fit of the fluctuation-induced forces versus distance for low stretching.
They were adjusted with the model of Eq. \ref{eq:f_ln_colloids},
by Lehle et al.\cite{Lehle_06}. Panel (a) corresponds to the case
$L^{*}=0.4$ and $T=5.05\varepsilon/k_{B}$, while panel (b) corresponds
to $L^{*}=0.5$ and $T=15.0\varepsilon/k_{B}$.\label{fig:ln_fit_F_vs_d}}
\end{figure}

We fitted Eq. \ref{eq:f_ln_colloids} with a multiplicative constant
$A$ and $c_{1}\equiv1/r_{0}$ as fitting parameters, and found a
very good agreement for the smaller chain stretchings ($L^{*}<0.65$)
at different temperatures. Two examples are provided in Fig. \ref{fig:ln_fit_F_vs_d},
for chain lengths of $L^{*}=0.4$ (panel (a)) and $L^{*}=0.5$ (panel
(b)). The lateral extension of the rings, that can be regarded as
the colloid radius $r_{0}$, is small as compared to typical ring
distance in the simulations, fulfilling the conditions for these limiting
case. The fitting parameter are around $A_{0}\simeq4$ for the multiplicative
constant of the force and $r_{0}\simeq0.36$ (with $r_{0}=0.14$ and
$r_{0}=0.55$ as the minimum and maximum values for all the fits).
This mean value seems a bit smaller than the effective ring width
(in its part exposed to the chain), but it is indeed in the order
of magnitude of effective ring width. 

\begin{figure}[t]
\begin{centering}
\includegraphics[clip,width=0.98\columnwidth]{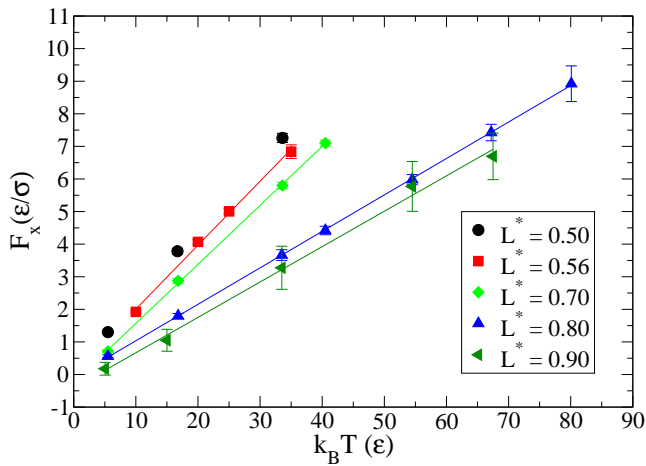} 
\par\end{centering}
\caption{\label{fig:Force-rings-vs-T-d=00003D6}Force between rings $F_{x}$
as a function of temperature for different chain lengths $L^{*}$.
In all the cases the rings were fixed at distance $d=6\sigma.$ The
lines come from a linear fit of the data point.}
\end{figure}

We also studied the dependence of the fluctuation-induced force with
the temperature. Fig. \ref{fig:Force-rings-vs-T-d=00003D6} shows
the fluctuation-induced force between rings located at a distance
of $d=6\sigma$ for different chain lengths. The curves present a
linear behavior. This is expected, because the fluctuation-induced
interactions are proportional to the driving energy of fluctuations,
$k_{B}T$ in our case\cite{Golestanian_96}. It is observed also a
smaller slope of the curves for higher chain stretching. This fact
can be rationalized by considering two facts. On one hand, the nonlinearity
of the bond potential make the bonds effectively stiffer for higher
stretching. The rate of increase of mode amplitudes, upon increase
in temperature, is therefore lower for more stretched chains. On the
other hand, the hindering of modes of higher amplitude (low $q$)
seems to be the most important contribution to the fluctuation induced
force. This can be observed in Fig. 13 ($d\gtrsim6\sigma)$, where
at equal temperature, the force is higher for shorter chains, i. e.
for chains of higher low-amplitude modes (see also Fig. 5). These
two facts may give rise to the smaller dependence with temperature
of more stretched chains observed in Fig. 15.

\begin{figure}[t]
\begin{centering}
\includegraphics[clip,width=0.98\columnwidth]{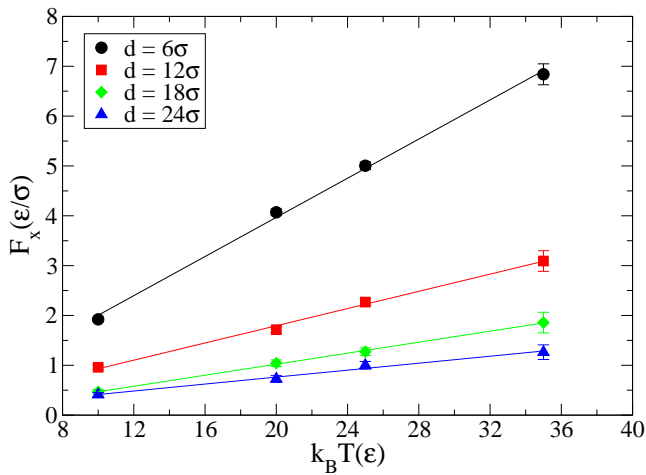}
\par\end{centering}
\caption{\foreignlanguage{spanish}{\label{fig:Fx_vs_temp_L*=00003D0.56}\foreignlanguage{american}{$F_{x}$
versus temperature for different ring distances. All the cases correspond
to $L^{*}=0.56$ and the number of beads in the chain is $N=1024$.
The lines are linear fits of the data for each case.}}}
\end{figure}

Figure \ref{fig:Fx_vs_temp_L*=00003D0.56} shows the force at different
ring distances as a function of temperature for a fixed chain stretching
of $L^{*}=0.56$. A linear dependence is also observed in this rather
narrow temperature interval. This indicates that the change of the
force is dominated by the number of modes which is the same at larger
temperatures, increasing the modes' amplitudes. In thermodynamic framework,
the term $T\Delta S$ of the free energy, depends on temperature basically
only through the explicit $T$ dependence and not in $\Delta S$.

Finally, we discuss in more detail the oscillations of the fluctuation-induced
force versus distance at very high stretching observed in Fig. \ref{fig:f_vs_d_different-stretching}.
In Fig. \ref{fig:Fprom_vs_Ns} we show the force versus ring distance
for $L^{*}=0.8$ for different chain lengths (Panel (a)) and different
temperatures (Panel (b)). A clear modulation of the force shows up
for short distances up to $d\simeq16\sigma$. This was observed in
the high stretching limit $L^{*}\gtrsim0.8$ for all the studied cases.
We recall that the maximum physical length that the FENE chain can
have is $L^{*}=1$, quite close to these cases. The beads are rather
separated and the center of the bonds have a clearly reduced excluded
volume, as compared to positions closer to the center of the beads.
In these conditions, for a given ring distance, it makes a significative
difference if the rings are located very close to a bead (see inset
(i) in Fig. \ref{fig:Fprom_vs_Ns}) or in the central region of the
bond (see inset (ii) in Fig. \ref{fig:Fprom_vs_Ns}). In the second
case, the chain can have fluctuations of higher amplitude inside the
ring, giving rise to a lower fluctuation-induced force. This is also
facilitated by the fact that the bond length lateral fluctuation is
rather low due to the high stretching of the polymer and the high
internal energy of the bond. We confirmed this by scaling the ring
distance with the mean bond distance $\langle l_{bond}\rangle$ for
each case. Plotting $F_{x}$ vs $d/\langle l_{bond}\rangle$ we observe
a period for the modulation of one bond length (not shown). The modulation
smudges at longer ring distances, where lateral displacements of the
beads close to the rings are much larger. 

While these stretching values are rather extreme and might be not
feasible experimentally, it draws the attention on the fact that the
structure of the bond and the details of the polymer structure at
the chemical level, could have a role in the modulation of fluctuation-induced
forces in experiments.

\begin{center}
\begin{figure}[t]
\begin{centering}
\begin{tabular}{c}
(a)\includegraphics[clip,width=0.98\columnwidth]{./comparo_en_funcion_del_N_1\lyxdot 14ln_33\lyxdot 6temp}\tabularnewline
(b)\includegraphics[clip,width=0.98\columnwidth]{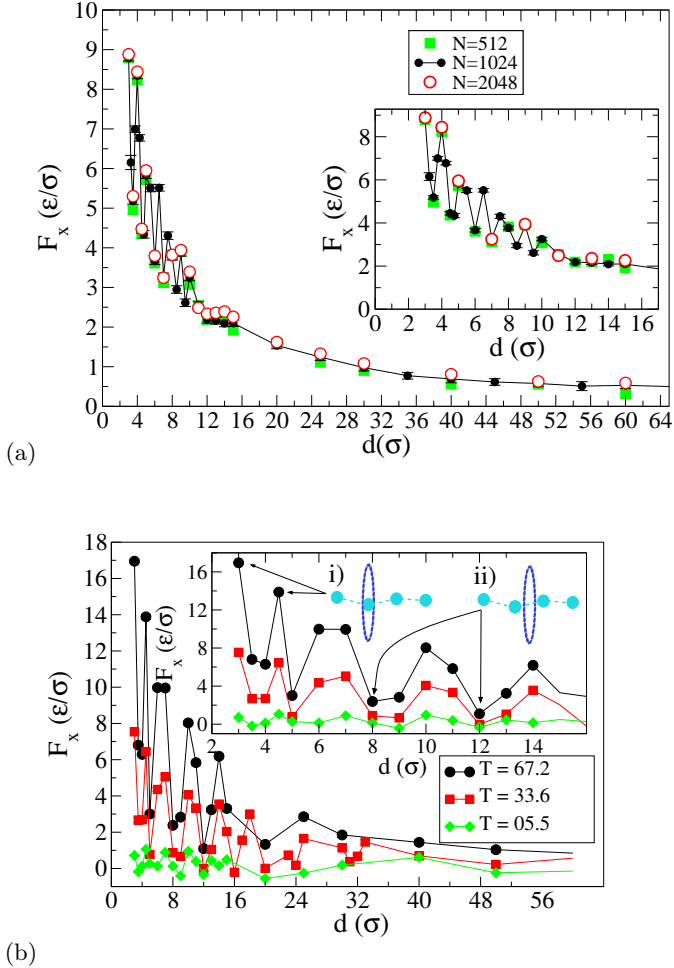}\tabularnewline
\end{tabular}
\par\end{centering}
\caption{Panel (a):\label{fig:Fprom_vs_Ns} Force versus distance for different
number of beads in the chain at the same chain stretching $L^{*}=0.80$.
Panel (b): $F_{x}$ as a function of distance for $L^{*}=0.8$ at
different temperatures. The insets in both panels show a detailed
view of $F_{x}$ for short distances. }
\end{figure}
\par\end{center}

\section{Discussion and Conclusions \label{sec:Conclusions}}

In this work, we studied the fluctuation properties of a polymer chain
under tension and the characteristics of the fluctuation-induced forces
between two ring molecules threaded around the polymer and fixed in the
space. In this way, the rings alter the natural fluctuations of a
stretched chain giving rise to the observed Casimir-like forces. The
system could be considered as a model of a pseudo-rotaxane under tension,
which could be studied experimentally as an isolated entity, or as
a component of a slide-ring material under tension. 

The unconstrained chain under tension presents a fluctuation spectrum
compatible with a $\sim1/q^{2}$ law for a wide range of studied chain
stretchings, which deviates from this behavior for lower chain lengths.
Interestingly, the behavior at high stretching is similar to a capillary
wave spectrum of a liquid-vapor interface or that of two immiscible
liquids, in spite of the non-linearity of the chain connectivity. 

We observed an attractive fluctuation-induced force between the rings
for all the studied cases. We characterize these forces as function
of chain stretching, temperature and ring radius, i.e. the properties
of the physical constraint imposed to the chain. For higher stretching,
we found a dependence of $\sim1/d$ of the fluctuation-induced force,
which is similar to that found in the context of a harmonic approximation
for classical electromagnetic fields at thermal equilibrium\cite{Boyer_03c},
and 2D Ising model of a pinned magnetic interface\cite{Abraham_07}.
We observed a linear dependence of the force with the temperature,
which is expected from an entropy-driven force of this type, if the
entropy differences between constraint and unconstrained chains are
not dependent on temperature. 

At lower stretchings, in which not all the degrees of freedom of the
chain are vibrating, we observe a deviation from this limiting behavior.
In this case the force vs. distance is adjusted very well with a dependence
$\sim1/d\ln(c_{1}d)$. This behavior was also found theoretically
in exact results for interfaces of 3D Ising systems\cite{Abraham_07} and for
colloids pinned in liquid-liquid interfaces\cite{Lehle_06} We also
characterized fluctuations of the force at very high stretchings,
coming from the discreteness of the beads and variations of the excluded
volume of the chain along the bonds. 

In addition to the results found for polymers under tension, we consider
the system as a relatively simple model to study fluctuation-induced
forces in different contexts. We are planning to continue this work,
by studying the dynamics of rings allowed to move in the direction
of the chain and also considering semi-flexible polymers with local
flexural rigidity, reminiscent of biofilaments. For this case, there
is an increment in the length of bond correlations, which could be
very interesting. We are also planning to study aggregation of ring
molecules threaded in the chain, in a system closer to a polyrotaxane.
Finally, we think that the direct measurement of these forces with
optical tweezers is plausible, or via a potential of mean force in
fluorescence experiments\cite{Bustamante_08b,Goldman_15}. 
\begin{acknowledgments}
C. P. warmly thanks M. Müller for enjoyable and fruitful discussions
along the development of this work. A. de Virgillis, with whom we
exchanged many ideas on the system studied here, is also gratefully
acknowledged. C. P. also thanks encouraging discussions with K. Binder
and T. Kreer in the early stages of this work. This research was supported
by CONICET (PIP 112 201101004 64), MINCYT (PICT-E 2014, PICT 1887-2011)
and CNEA (INN-CNEA 2011) . R. L. thanks UTN for the ph D scholarship 
(UTN Rectorado Res. 1351/14) which supports her work. 
\end{acknowledgments}

%\bibliographystyle{apsrev_no_url}
%\bibliography{bibtex}

\end{document}